\documentclass[prl,twocolumn,showpacs,superscriptaddress]{revtex4}

\usepackage{amsmath,amssymb}
\usepackage{verbatim}
\usepackage{graphicx}
\usepackage{hyperref}
\usepackage{color}

\DeclareFontFamily{OT1}{rsfs}{}
\DeclareFontShape{OT1}{rsfs}{m}{n}{ <-7> rsfs5 <7-10> rsfs7 <10->rsfs10}{}
\DeclareMathAlphabet{\mycal}{OT1}{rsfs}{m}{n}

\newcommand{\rmd}{\mathrm{d}}
\newcommand{\rmi}{\mathrm{i}}

\newcommand{\be}[1]{ \begin{equation}\label{#1} }
\newcommand{\ee}{\end{equation}}
\newcommand{\bea}[1]{\begin{eqnarray}\label{#1} }
\newcommand{\eea}{\end{eqnarray}}
\begin{document}

\title{Three-Dimensional Extended Bargmann Supergravity}

\author{Eric Bergshoeff}
\email{e.a.bergshoeff@rug.nl}
\affiliation{Van Swinderen Institute for Particle Physics and Gravity, University of Groningen, Nijenborgh 4, 9747 AG Groningen, The Netherlands}

\author{Jan Rosseel}
\email{rosseel@itp.unibe.ch}
\affiliation{Albert Einstein Center for Fundamental Physics, University of Bern, Sidlerstrasse 5, 3012 Bern, Switzerland}

\date{\today}

\preprint{TUW--13--09}

\begin{abstract}
We show that three-dimensional General Relativity, augmented with two vector fields,
allows for a non-relativistic limit, different from the standard limit leading  to Newtonian gravity, that results into a well-defined action which is of the Chern-Simons type. We show that this three-dimensional `Extended Bargmann Gravity', after coupling to matter, leads to equations of motion allowing a wider class of background geometries than the ones that one encounters in Newtonian gravity.
We give the supersymmetric generalization of these results and point out an important application in the context of calculating partition functions of non-relativistic field theories using localization techniques.

\end{abstract}

\pacs{04.60.Kz, 11.15.Yc, 11.25.Tq}

\maketitle

\section{Introduction}

Even though supergravity is by now four decades old \cite{Freedman:1976xh}, non-relativistic versions of it have only been constructed  recently, in three space-time dimensions only \cite{Andringa:2013mma,Bergshoeff:2015ija}. This is surprising in view of the fact that non-relativistic supergravity is a natural starting point to extend to the non-relativistic realm the numerous applications of supergravity in relativistic field theory and holography. Constructing non-relativistic supergravity theories is non-trivial, due to the requirement that they obey mass (or particle number) conservation. This can be achieved  by introducing a gauge field that  couples to the mass or particle number current. The difficulty in constructing non-relativistic  supergravity lies in the fact that mass conservation corresponds to a space-time symmetry, that does not have a  relativistic analogue. As a consequence, the gauge field coupling to the mass currents is a part of the non-relativistic supergravity multiplet that  does not have an obvious counterpart in the relativistic multiplet \footnote{
We note  that particular non-relativistic gravity theories coupled to matter can  be obtained from higher-dimensional relativistic gravity via null reduction \cite{Julia:1994bs,Bleeken:2015ykr}. It would be interesting to extend this procedure to supergravity.}.

The constructions of \cite{Andringa:2013mma,Bergshoeff:2015ija} incorporate mass conservation by using a differential geometric framework for non-relativistic space-times, that is called Newton-Cartan (NC) geometry \cite{Cartan1,Cartan2,Dombrowski,Kunzle1972,Misner:1974qy,Kuchar:1980tw,Duval:1983pb,Duval:1984cj,Andringa:2010it,Bekaert:2014bwa,Bekaert:2015xua}. In one formulation of NC geometry, one not only provides a space-time with  a time-like and spatial metric, but also with a gauge field that couples to the mass current. Recently, NC geometry has found unexpected applications
in holographic approaches to describe strongly coupled condensed matter systems \cite{Son:2008ye,Kachru:2008yh,Balasubramanian:2008dm,Hartnoll:2009sz,Christensen:2013lma,
Christensen:2013rfa,Hartong:2014oma,Hartong:2014pma,Hartong:2015wxa} as well as
in the construction of non-relativistic effective field theories describing condensed matter models such
as the Fractional Quantum Hall Effect, chiral superfluids and simple fluids \cite{Son:2005rv,Hoyos:2011ez,Son:2013rqa,Abanov:2014ula,Gromov:2014vla,Geracie:2014nka}.
NC geometry has recently also led to new insights into the structure of Ho\v rava-Lifshitz gravity \cite{Hartong:2015zia}. Many of these applications involve matter coupled to arbitrary NC backgrounds.

The theories of \cite{Andringa:2013mma,Bergshoeff:2015ija} are supersymmetric extensions of Newton-Cartan gravity \cite{Cartan1,Cartan2}, i.e.~Cartan's reformulation of Newtonian gravity in arbitrary coordinate frames, akin to General Relativity. These NC supergravity theories are, however, restrictive, in that they do not admit space-times with non-trivial spatial curvature as solutions. Indeed, the field equations of NC gravity state that the purely time-like component of the Ricci tensor is proportional to the mass density, while all other components are zero \footnote{These field equations constitute a covariant version of the Poisson equation. In dimensions $D \geq 4$, they can  be derived as a non-relativistic limit of the Einstein equations, where the limit is taken as explained in \cite{Bergshoeff:2015uaa}. In three dimensions, one finds that in this limit the mass density no longer sources the time-like component of the Ricci tensor. One can still postulate field equations in which this component of the Ricci tensor is proportional to the mass density, but  such equations cannot be viewed as a non-relativistic limit.}. Backgrounds with non-trivial spatial curvature are thus not allowed and this is equally well true in NC supergravity, where this  holds for supersymmetric backgrounds in particular \cite{Knodel:2015byb}. A further complication  is that at present no satisfactory action principle for NC supergravity exists.

In this letter, we will show that, at least in three dimensions, there exists an alternative  non-relativistic supergravity theory, that takes away  the above mentioned restrictions. This supergravity is the supersymmetric extension  of a theory that we will call `Extended Bargmann Gravity' (EBG) and that was first considered in \cite{Papageorgiou:2009zc}. EBG is a Chern-Simons theory for a central extension of the so-called Bargmann algebra that consists of the isometries of non-relativistic flat space-times. This is similar to how three-dimensional General Relativity 
can be viewed as a Chern-Simons theory for the Poincar\'e 
algebra \cite{Witten:1988hc,Achucarro:1987vz}. Here, we will first revisit EBG and show how it can be obtained as a non-relativistic limit of a suitable generalization of the Einstein-Hilbert action. This derivation, along with the existence of an action, makes it easy to include matter, by building on previous results obtained in \cite{Bergshoeff:2015sic}. By studying matter couplings we will show that in EBG it is possible to have non-trivial spatial curvature when matter is present, analogously to what happens in three-dimensional General Relativity \cite{Deser:1983tn}. Next, we will show that EBG has a supersymmetric extension, by giving a particular example of a superalgebra with invariant supertrace that contains the gauge algebra of EBG and by constructing the associated super-Chern-Simons action.

Super-EBG is particularly relevant in view of recent applications of relativistic matter actions coupled to off-shell supergravity as a tool to construct supersymmetric field theories on curved backgrounds \cite{Festuccia:2011ws}, whose non-perturbative behaviour can be studied using localization techniques \cite{Pestun:2007rz,Marino:2012zq}. Since super-EBG  allows for  supersymmetric curved backgrounds, it can be used as a starting point to extend these applications to non-relativistic field theories.

\section{Extended Bargmann Gravity}

The space-time symmetry algebra of EBG consists of the generators of time translations $H$, spatial translations $P_a$ (with $a=1,2$), spatial rotations $J$, Galilean boosts $G_a$, a central charge $M$, corresponding to particle mass as well as a second central charge $S$ \cite{Levy-Leblond:1971,Grigore:1993fz,Bose:1994sj,Jackiw:2000tz}.
The generators $H$, $P_a$, $J$, $G_a$ and $M$ form the so-called Bargmann algebra and the inclusion of $S$ leads to what we will  refer to as the `extended Bargmann algebra' whose non-zero commutation relations are given by
\begin{alignat}{2}
  \label{eq:Bargmann2cc}
  \left[H, G_a\right] &= -\epsilon_{ab} P_b \,, \quad & \left[J, G_a\right] &= -\epsilon_{ab} G_b \,, \nonumber \\  \left[J, P_a\right] &= -\epsilon_{ab} P_b \,, \quad & \left[G_a, G_b\right] &= \epsilon_{ab} S \,,  \nonumber \\
  \left[G_a, P_b\right] &= \epsilon_{ab} M \,.
\end{alignat}
Unlike the Bargmann algebra, the extended Bargmann algebra can be equipped with a non-degenerate, invariant bilinear form or `trace', given by \cite{Papageorgiou:2009zc}
\begin{equation}
  \label{eq:bilformnonrel}
  <G_a, P_b> = \delta_{ab} \,, \quad <H,S> =  <M,J> = -1 \,.
\end{equation}
The action of EBG is  given by the Chern-Simons action for the gauge algebra (\ref{eq:Bargmann2cc})
\begin{equation}
  \label{eq:CSaction}
  S = \frac{k}{4 \pi} \int <A \wedge dA + \frac23 A \wedge A \wedge A > \,,
\end{equation}
where $k$ is the Chern-Simons coupling constant and the gauge field $A = A_\mu \rmd x^\mu$ is given by
\begin{equation}
  \label{eq:Aexp}
  A_\mu = \tau_\mu \, H + e_\mu{}^a \, P_a + \omega_\mu\, J + \omega_\mu{}^a \, G_a + m_\mu\, M + s_\mu\, S \,.
\end{equation}
Explicitly, one finds the following action for EBG \cite{Papageorgiou:2009zc}\footnote{We thank Diego Hofman for sharing with us the observation that a particular truncation of
this action is related to the lower-spin gravity action introduced in \cite{Hofman:2014loa}.}
\begin{align}
  \label{eq:CSactionnonrelbos}
  S &= \frac{k}{4 \pi} \int \, \rmd^3 x \, \Big( \epsilon^{\mu\nu\rho} e_\mu{}^a R_{\nu\rho}(G_a) - \epsilon^{\mu\nu\rho} m_\mu R_{\nu\rho}(J) \nonumber \\ & \qquad \qquad - \epsilon^{\mu\nu\rho} \tau_\mu R_{\nu\rho}(S) \Big) \,,
\end{align}
where here and in the following we have used the curvatures
\begin{align}
  \label{eq:covcurvsBargmann}
  R_{\mu\nu}(H) &= 2 \partial_{[\mu} \tau_{\nu]} \,, \nonumber \\
  R_{\mu\nu}(P^a) &= 2 \partial_{[\mu} e_{\nu]}{}^a + 2 \epsilon^{ab} \omega_{[\mu} e_{\nu]b} - 2 \epsilon^{ab} \omega_{[\mu b} \tau_{\nu]} \,, \nonumber \\
  R_{\mu\nu}(J) &= 2 \partial_{[\mu} \omega_{\nu]} \,, \nonumber \\
  R_{\mu\nu}(G^a) &= 2 \partial_{[\mu} \omega_{\nu]}{}^a + 2 \epsilon^{ab} \omega_{[\mu} \omega_{\nu] b} \,, \nonumber \\
  R_{\mu\nu}(M) &= 2 \partial_{[\mu} m_{\nu]} + 2 \epsilon^{ab} \omega_{[\mu a} e_{\nu] b} \,, \nonumber \\
  R_{\mu\nu}(S) &= 2 \partial_{[\mu} s_{\nu]} + \epsilon^{ab} \omega_{[\mu a} \omega_{\nu] b} \,.
\end{align}
These curvatures are covariant with respect to the local $H$, $P_a$, $J$, $G_a$, $M$ and $S$ transformations of $\tau_\mu$, $e_\mu{}^a$, $m_\mu$, $\omega_\mu$, $\omega_\mu{}^a$ and $s_\mu$, that are found from the gauge algebra (\ref{eq:Bargmann2cc}) following the usual rules of gauge theory.
Note that the fields $\tau_\mu$, $e_\mu{}^a$, $\omega_\mu$, $\omega_\mu{}^a$ and $m_\mu$ also appear in the formulation of NC gravity obtained by gauging the Bargmann algebra. As in that case, $\tau_\mu$ and $e_\mu{}^a$ can be interpreted as Vielbeine \footnote{Even though these Vielbeine are not invertible, one can define projective inverses $\tau^\mu$, $e^\mu{}_a$ via the relations $\tau^\mu \tau_\mu = 1$, $\tau^\mu e_\mu{}^a = 0$, $\tau_\mu e^\mu{}_a = 0$, $e^{\mu}{}_a e_\mu{}^b = \delta^b_a$, $e^\mu{}_a e_\nu{}^a = \delta^\mu_\nu - \tau^\mu \tau_\nu$.
} for two degenerate time-like and spatial metrics, respectively.
The field $s_\mu$ is not present in NC gravity and is specific to EBG. We note that the equations of motion for $s_\mu$, $\omega_\mu$ and $\omega_\mu{}^a$ lead to the curvature constraints
\begin{align}
  \label{eq:eomssomegas}
  R_{\mu\nu}(H) = 0 \,, \quad R_{\mu\nu}(P^a) = 0 \,, \quad R_{\mu\nu}(M) = 0 \,,
\end{align}
that are usually imposed by hand in NC gravity.
As in NC gravity, these equations imply that EBG is defined on non-relativistic space-times with torsionless NC geometry. The first equation implies that the space-time can be foliated in an absolute time direction, while the last two equations can be used to express $\omega_\mu$ and $\omega_\mu{}^a$ in terms of $\tau_\mu$, $e_\mu{}^a$ and $m_\mu$. Following NC gravity, $\omega_\mu$ and $\omega_\mu{}^a$ can be seen as appropriate non-relativistic versions of the relativistic spin connection.

The EBG action (\ref{eq:CSactionnonrelbos}) can be obtained as the non-relativistic limit of a suitable extension of the three-dimensional Einstein-Hilbert action. In order to show this, we extend the procedure developed in \cite{Bergshoeff:2015uaa}, that allows one  to obtain the equations of motion of NC gravity from Einstein's equations. 
As a starting point, we take the following Einstein-Hilbert action for the relativistic Vielbein $E_\mu{}^A$ and spin connection $\Omega_\mu{}^{AB}$, written as a Chern-Simons action, plus a Chern-Simons action for two abelian gauge fields $Z_{1 \mu}$ and $Z_{2 \mu}$:
\begin{equation}
  \label{eq:CSactionrel}
  S = \frac{k \omega}{4\pi} \int \, \rmd^3 x \,  \Big(\epsilon^{\mu\nu\rho}\, E_\mu{}^A \, R_{\nu\rho}(J_A) + 2\,\epsilon^{\mu\nu\rho}\, Z_{1 \mu}\, \partial_\nu Z_{2 \rho} \Big) \,,
\end{equation}
where the Riemann tensor $R_{\mu\nu}(J^A)$ reads
\begin{equation}
  \label{eq:defRrel}
  R_{\mu\nu}(J^A) = 2 \partial_{[\mu} \Omega_{\nu]}{}^A -  \epsilon^{ABC} \Omega_{[\mu B} \Omega_{\nu] C} \,.
\end{equation}
 Extending the particle-limit procedure of \cite{Bergshoeff:2015uaa}, mimicking the In\"on\"u-Wigner contraction of the underlying Poincar\'e $\otimes$ U(1)${}^2$ gauge algebra, we express  the relativistic gauge fields $E_\mu{}^A$, $\Omega_\mu{}^A$, $Z_{1\mu}$, $Z_{2\mu}$ in
 terms of the  non-relativistic
 fields $\tau_\mu,e_\mu{}^a, \omega_\mu{}^a, \omega_\mu, m_\mu, s_\mu$ as follows:
\begin{alignat}{2}
  \label{eq:fieldcontr}
  E_\mu{}^0 &= \omega \, \tau_\mu + \frac{1}{2\omega}\, m_\mu\,, \qquad \qquad & Z_{1\mu} &= \omega \, \tau_\mu - \frac{1}{2\omega} \, m_\mu \,, \nonumber \\
  \Omega_\mu{}^0 &= \omega_\mu + \frac{1}{2\omega^2} \, s_\mu \,, \qquad \qquad & Z_{2\mu} &= \omega_\mu - \frac{1}{2\omega^2} \, s_\mu \,, \nonumber \\
  E_\mu{}^a &= e_\mu{}^a \,, \qquad \qquad & \Omega_\mu{}^a &= \frac{1}{\omega}\, \omega_\mu{}^a \,.
\end{alignat}
 Using these expressions  in the action (\ref{eq:CSactionrel}) and taking the limit $\omega \rightarrow \infty$ \footnote{This limit is well-defined. The term $-2 \, \epsilon^{\mu\nu\rho} E_\mu{}^0 \, \partial_\nu \Omega_\rho{}^0$ leads to a potentially diverging $-2\, \omega^2\, \epsilon^{\mu\nu\rho} \tau_\mu \, \partial_\nu \omega_\rho$ term, but this term gets cancelled by a contribution coming from the term $2\, \epsilon^{\mu\nu\rho} \, Z_{1\mu} \, \partial_\nu Z_{2\rho}$
 that we added to the Einstein-Hilbert action.} it is straightforward to show that  the EBG action (\ref{eq:CSactionnonrelbos}) is obtained.

In the next section we will show that EBG and NC gravity are different theories by comparing the coupling to matter.

\section{Matter coupling}

Introducing matter couplings in EBG can be done by adding one of the matter Lagrangians on arbitrary torsionless NC backgrounds constructed in 
\cite{Bergshoeff:2015sic}:
\begin{align}
  \label{eq:mattcouplaction}
  S &= \frac{k}{4\pi} \int \rmd^3 x \, \Big( \epsilon^{\mu\nu\rho} e_\mu{}^a R_{\nu\rho}(G_a) - \epsilon^{\mu\nu\rho} \tau_\mu R_{\nu\rho}(S) \nonumber \\ &- \epsilon^{\mu\nu\rho} m_\mu R_{\nu\rho}(J) \Big) + \int \rmd^3 x\, e \, \mathcal{L}_{\mathrm{m}} \,.
\end{align}
Here $e = \mathrm{det}(\tau_\mu, e_\mu{}^a)$ denotes the volume element.
Since any matter couplings to the $s_\mu$ gauge field change the foliation constraint $R_{\mu\nu}(H) = 0$, we do not consider such couplings so that we can stay within the framework of torsionless NC geometry.

Defining the energy current $t^\mu$, the momentum current $t^\mu{}_a$ and the particle number current $j^\mu$ by
\begin{align}
  \label{eq:defcurrents}
  t^\mu &= \frac{1}{e}\frac{\delta}{\delta \tau_\mu} \left(e \mathcal{L}_{\mathrm{m}} \right) \,, \qquad t^\mu{}_a = \frac{1}{e}\frac{\delta}{\delta e_\mu{}^a} \left(e \mathcal{L}_{\mathrm{m}} \right) \,, \nonumber \\  j^\mu &=\frac{1}{e} \frac{\delta}{\delta m_\mu} \left(e \mathcal{L}_{\mathrm{m}} \right) \,,
\end{align}
the equations of motion stemming from the action (\ref{eq:mattcouplaction}) take the form
\begin{alignat}{2}
  \label{eq:eomswithmatter}
  e^{-1} \epsilon^{\mu\nu\rho} R_{\nu\rho}(S) &= \frac{4\pi}{k} t^\mu \,, \ \ \ &
  e^{-1} \epsilon^{\mu\nu\rho} R_{\nu\rho}(J) &= \frac{4\pi}{k} j^\mu \,, \nonumber \\
  e^{-1} \epsilon^{\mu\nu\rho} R_{\nu\rho}(G_a) &= -\frac{4\pi}{k} t^\mu{}_a \,.
\end{alignat}
Since the curvatures in these equations obey Bianchi identities, the currents  obey various identities for consistency. We distinguish between Bianchi identities `of the first kind' and `of the second kind'. The identities of the first kind follow from the fact that the equations
\begin{equation} \label{eq:convandunconv}
  R_{\mu\nu}(P^a) = 0 \,, \qquad R_{\mu\nu}(M) = 0 \,
\end{equation}
are identically satisfied, once one views the spin connections $\omega_\mu$ and $\omega_\mu{}^a$ as dependent on $\tau_\mu$, $e_\mu{}^a$ and $m_\mu$. Substituting  equations \eqref{eq:convandunconv}  into  the  Bianchi identities $D_{[\mu} R_{\nu\rho]}(P^a) = 0$ and $D_{[\mu} R_{\nu\rho]}(M) = 0$  leads to the following Bianchi identities of the first kind:
\begin{equation}
  \label{eq:firstkindBianchis}
  R_{[\mu\nu}(J)\, e_{\rho]}{}^a = R_{[\mu\nu}(G^a)\, \tau_{\rho]} \,, \qquad \epsilon^{ab} R_{[\mu\nu}(G_a) \, e_{\rho]b} = 0 \,.
\end{equation}
The remaining Bianchi identities, called  of the second kind, are not algebraic in the curvatures and  are given by
\begin{equation}
  \label{eq:secondkindBianchis}
  D_{[\mu} R_{\nu\rho]}(G^a) = 0 \,, \quad \partial_{[\mu} R_{\nu\rho]}(J) = 0 \,, \quad D_{[\mu} R_{\nu\rho]}(S) = 0 \,.
\end{equation}
Combining the equations of motion (\ref{eq:eomswithmatter}) with the Bianchi identities of the first kind (\ref{eq:firstkindBianchis})  leads to the following
algebraic consistency conditions
\begin{equation}
  \label{eq:conscondcurrentssymm}
  e_\mu{}^a j^\mu = - \tau_\mu  t^\mu{}_a\,, \qquad e_\mu{}^{[a} t^{|\mu|b]} = 0 \,.
\end{equation}
The Bianchi identities of the second kind on the other hand lead to the following current conservation conditions:
\begin{equation} \label{eq:currentconservation}
  D_\mu t^\mu = 0 \,, \qquad  D_\mu t^\mu{}_a = 0 \,, \qquad D_\mu j^\mu = 0 \,.
\end{equation}

The EBG equations of motion (\ref{eq:eomswithmatter}) are strikingly different from the NC gravity ones. 
Using the identity
\begin{equation}
  \label{eq:riemanndef}
  R^\sigma{}_{\rho\mu\nu} = \epsilon^{ab} R_{\mu\nu}(J) e^\sigma{}_a e_{\rho b} - \epsilon^{ab} R_{\mu\nu}(G_b) e^\sigma{}_a \tau_\rho 
\end{equation}
it follows from the equations of motion (\ref{eq:eomswithmatter}) that  the purely time-like component of the Ricci tensor $R_{\mu\nu} = R^\rho{}_{\mu\rho\nu}$ is given by $
  \tau^\mu \tau^\nu R_{\mu \nu} \propto e_{\mu}{}^a\, t^\mu{}_a$.
This is unlike NC gravity, where one rather has $\tau^\mu \tau^\nu R_{\mu\nu} \propto j^0$ \footnote{It is instructive to consider this difference in the example of a massive complex Schr\"odinger field $\Phi$, whose action can be found in \cite{Bergshoeff:2015sic}. In that case one finds for EBG that $\tau^\mu \tau^\nu R_{\mu \nu} = -\frac{2\pi \rmi}{k} \left(\Phi^* D_0 \Phi - \Phi D_0 \Phi^* \right)$ while the  case of NC gravity leads to  $\tau^\mu \tau^\nu R_{\mu \nu} \propto |\Phi|^2$.}. Furthermore, in NC gravity only this  purely time-like component of the Ricci tensor is non-zero.  This is in contrast with EBG where matter sources all components of the Riemann tensor.
As a result, three-dimensional EBG  admits backgrounds with non-trivial curvature whenever matter is present.

\section{Extended Bargmann Supergravity}

To construct a supersymmetric extension  of EBG one needs to find a superalgebra with a `supertrace' that contains the extended Bargmann algebra (\ref{eq:Bargmann2cc}) as a subalgebra.
 By trial and error we have found that the Bargmann algebra
(\ref{eq:Bargmann2cc}) can be extended, not necessarily uniquely,  with three fermionic generators $Q^+$, $Q^-$ and $R$ that are all Majorana spinors. The latter generator is reminiscent to the fermionic generator introduced in \cite{Green:1989nn,Bergshoeff:1989ax}.
Apart from the commutation relations given in (\ref{eq:Bargmann2cc}), this superalgebra has the following non-zero (anti-)\-commutation relations:
\begin{alignat}{2}
  \label{eq:superBargmann2cc}
 & [J, Q^\pm] = -\frac12 \gamma_0 Q^\pm \,,  \ & & [J, R] = -\frac12 \gamma_0 R \,, \nonumber \\
 & [G_a, Q^+] = -\frac12 \gamma_a Q^- \,,  \ & & [G_a, Q^-] = -\frac12 \gamma_a R \,, \nonumber \\
 & [S, Q^+] = - \frac12 \gamma_0 R \,, \  & & \{Q^+_{\alpha},Q^+_{\beta}\} = (\gamma_0 C^{-1})_{\alpha\beta} H \,, \nonumber \\
 & \{Q^+_{\alpha},Q^-_{\beta}\} = -(\gamma^a C^{-1})_{\alpha\beta} P_a \,, \  & & \{Q^-_{\alpha},Q^-_{\beta}\} = (\gamma_0 C^{-1})_{\alpha\beta} M \,, \nonumber \\
 & \{Q^+_{\alpha}, R_\beta\} = (\gamma_0 C^{-1})_{\alpha\beta} M \,.
\end{alignat}
The invariant supertrace on this algebra is given by (\ref{eq:bilformnonrel}), extended with
\begin{equation}
  \label{eq:supertrace}
  <Q^+_\alpha, R_\beta> = 2 (C^{-1})_{\alpha\beta} \,, \quad <Q^-_\alpha, Q^-_\beta> = 2 (C^{-1})_{\alpha\beta} \,.
\end{equation}
Introducing the gauge field
\begin{align}
  \label{eq:Asuper}
  A_\mu &= \tau_\mu H + e_{\mu}{}^a P_a + \omega_\mu J + \omega_\mu{}^a G_a + m_\mu M + s_\mu S \nonumber \\ & + \bar{\psi}^+_\mu Q^+ + \bar{\psi}^-_\mu Q^- + \bar{\rho}_\mu R \,,
\end{align}
the Chern-Simons action for the superalgebra (\ref{eq:superBargmann2cc}) is  given by
\begin{align}
  \label{eq:CSactionnonrelsuper}
  S &= \frac{k}{4 \pi} \int \, \rmd^3 x \, \Big( \epsilon^{\mu\nu\rho} e_\mu{}^a R_{\nu\rho}(G_a) - \epsilon^{\mu\nu\rho} m_\mu R_{\nu\rho}(J) \nonumber \\ & \qquad \qquad - \epsilon^{\mu\nu\rho} \tau_\mu R_{\nu\rho}(S) + \epsilon^{\mu\nu\rho} \bar{\psi}^+_\mu \hat{\rho}_{\nu\rho} + \epsilon^{\mu\nu\rho} \bar{\rho}_\mu \hat{\psi}^+_{\nu\rho} \nonumber \\ & \qquad \qquad + \epsilon^{\mu\nu\rho} \bar{\psi}^-_\mu \hat{\psi}^-_{\nu\rho}\Big) \,,
\end{align}
where the supercovariant curvatures are given by
\begin{align}
  \label{eq:supcovcurvs}
& \hat{\psi}^+_{\mu\nu} = 2 \partial_{[\mu} \psi^+_{\nu]} + \omega_{[\mu} \gamma_0 \psi_{\nu]}^+ \,, \nonumber \\
&  \hat{\psi}^-_{\mu\nu} = 2 \partial_{[\mu} \psi^-_{\nu]} + \omega_{[\mu} \gamma_0 \psi_{\nu]}^- + \omega_{[\mu}{}^a \gamma_a \psi_{\nu]}^+ \,, \nonumber \\
& \hat{\rho}_{\mu\nu} = 2 \partial_{[\mu} \rho_{\nu]} + \omega_{[\mu} \gamma_0 \rho_{\nu]} + \omega_{[\mu}{}^a \gamma_a \psi_{\nu]}^- + s_{[\mu} \gamma_0 \psi_{\nu]}^+ \,.
\end{align}
These curvatures 
transform covariantly with respect to the  supersymmetry transformation rules
\begin{align}
  \label{eq:susytrafos}
  \delta \tau_\mu &= -\bar{\epsilon}^+ \gamma_0 \psi^+_\mu \,, \nonumber \\
  \delta e_\mu{}^a &= \bar{\epsilon}^+ \gamma^a \psi^-_\mu + \bar{\epsilon}^- \gamma^a \psi^+_\mu \,, \nonumber \\
  \delta m_\mu &= -\bar{\epsilon}^- \gamma_0 \psi_\mu^- - \bar{\epsilon}^+ \gamma_0 \rho_\mu - \bar{\eta} \gamma_0 \psi_\mu^+ \,, \nonumber \\
  \delta \psi_\mu^+ &= \partial_\mu \epsilon^+ + \frac12 \omega_\mu \gamma_0 \epsilon^+ \,, \nonumber \\
  \delta \psi_\mu^- &= \partial_\mu \epsilon^- + \frac12 \omega_\mu \gamma_0 \epsilon^- + \frac12 \omega_\mu{}^a \gamma_a \epsilon^+  \,, \nonumber \\
  \delta \rho_\mu &= \partial_\mu \eta + \frac12 \omega_\mu \gamma_0 \eta + \frac12 \omega_\mu{}^a \gamma_a \epsilon^- + \frac12 s_\mu \gamma_0 \epsilon^+ \,,
\end{align}
where $\epsilon^\pm$ and $\eta$ are the parameters of the local $Q^\pm$ and $R$ transformations, respectively.

The action \eqref{eq:CSactionnonrelsuper} shows that there exists at least one   extended Bargmann supergravity theory in three spacetime dimensions
that is different from the NC supergravity theory constructed in \cite{Andringa:2013mma,Bergshoeff:2015ija}.

\section{Outlook}

The work presented here  serves as a starting point for various further studies. First of all, it would be interesting to see whether a cosmological constant can be included and whether extensions of the so-called super-Newton-Hooke algebra can be found that admit an invariant supertrace. Secondly, it would be interesting to see whether matter couplings exist that involve the vector field $s_\mu$ since this would automatically lead to a torsionfull NC geometry.

Extended Bargmann Gravity  will be useful to explore various aspects of holography in the non-relativistic limit.
Defining a solution space of EBG that obeys appropriate boundary conditions could serve as a representation space for a potentially infinite-dimensional asymptotic symmetry algebra. Since EBG is the non-relativistic analogue of Einstein gravity in flat space-time, such an asymptotic symmetry algebra would  be the non-relativistic version of the BMS algebra. As such, EBG can be used to   study aspects of flat space holography in the non-relativistic limit. Furthermore, it would be interesting to see whether a higher spin version of EBG corresponding to the non-relativistic limit of \cite{Afshar:2013vka} can be constructed.
This could be used to explore the non-relativistic limit of higher spin holography in flat space-times.

The superalgebra that we  presented in this letter has not been derived as an In\"on\"u-Wigner contraction of a relativistic superalgebra. It remains to be seen whether  this is possible and whether different superalgebra extensions of (\ref{eq:Bargmann2cc}) with  invariant supertrace exist. In order to construct matter-coupled supergravity extensions of EBG, it would be useful to study supermultiplet representations of the underlying  superalgebra. The supersymmetric extension of matter coupled EBG
will lead to new opportunities for constructing non-relativistic supersymmetry in  non-trivial backgrounds and as such will have important applications for the calculation of non-relativistic partition functions using  localization techniques \cite{Pestun:2007rz,Marino:2012zq}.

Upon the completion of this work we were informed by Jelle Hartong and Niels Obers that the EBG action \eqref{eq:CSactionnonrelbos}
can be identified as a particular kinetic term of three-dimensional Ho\v rava-Lifshitz gravity \cite{Hartong:2016yrf}.
Using this interpretation our results are relevant for the construction of the supersymmetric extension of three-dimensional Ho\v rava-Lifshitz gravity.

\section{Acknowledgements}

We are grateful to A. Bagchi, M. Blau, J. Gomis, D. Grumiller, D. Hofman, J. Hartong, N. Obers and T. Zojer for useful discussions. The work of JR was supported by the NCCR SwissMAP, funded by the Swiss National Science Foundation.

%\bibliography{NR3DsCS}


\end{document}